\documentclass[twocolumn,showpacs,preprintnumbers,amsmath,amssymb,superscriptaddress]{revtex4}
\usepackage{graphicx}
\usepackage{dcolumn}
\usepackage{bm}

\begin{document}

\title{Acoustic-phonon-based interaction between coplanar nano-circuits in magnetic field}

\author{M.G.~Prokudina}
\affiliation{Institute of Solid State Physics, Russian Academy of
Sciences, 142432 Chernogolovka, Russian Federation}
\author{V.S.~Khrapai}
\affiliation{Institute of Solid State Physics, Russian Academy of
Sciences, 142432 Chernogolovka, Russian Federation}
\author{S.~Ludwig}
\affiliation{Center for NanoScience and Department
f$\ddot{\text{u}}$r Physik,
Ludwig-Maximilians-Universit$\ddot{\text{a}}$t,
Geschwister-Scholl-Platz 1, D-80539 M$\ddot{\text{u}}$nchen,
Germany}
\author{J.P.~Kotthaus}
\affiliation{Center for NanoScience and Department
f$\ddot{\text{u}}$r Physik,
Ludwig-Maximilians-Universit$\ddot{\text{a}}$t,
Geschwister-Scholl-Platz 1, D-80539 M$\ddot{\text{u}}$nchen,
Germany}
\author{H.P.~Tranitz}
\affiliation{Institut f$\ddot{\text{u}}$r Experimentelle und
Angewandte Physik, Universit$\ddot{\text{a}}$t Regensburg, D-93040
Regensburg, Germany}
\author{W.~Wegscheider} \affiliation{Solid State Physics Laboratory, ETH Zurich, 8093 Zurich, Switzerland}

\begin{abstract}

We explore the acoustic phonon-based interaction between two
neighboring coplanar circuits containing semiconductor quantum
point contacts in a perpendicular magnetic field $B$. In a
drag-type experiment, a current flowing in one of the circuits
(unbiased) is measured in response to an external current in the
other. In moderate $B$ the sign of the induced current is
determined solely by the polarity of $B$. This indicates that the
spatial regions where the phonon emission/reabsorption is
efficient are controlled by magnetic field. The results are
interpreted in terms of non-equilibrium transport via skipping
orbits in two-dimensional electron system.

\end{abstract}

\maketitle

Non-equilibrium behavior in coupled nano-circuits in solid state
devices is often studied in drag-type experiments. That is, an
external current is sent through one (drive-) circuit and a
current (or voltage) induced is measured in a second neighboring
(detector-) circuit. Different interaction mechanisms include
Coulomb drag~\cite{zverev,yamamoto}, exchange of ultra
high-frequency photons~\cite{aguado,onac,gustavsson} or other
energy quanta~\cite{DQDratchet}, rectification~\cite{kamenev} and
coulombic or phononic backaction~\cite{gasser,schinner,harbusch}.

A typical size of a lateral nanostructure defined in a GaAs-based
two-dimensional electron system (2DES) is on the order of 1$\mu$m.
In comparison, the nanostructure transport current probes a
distribution function of carriers in the 2DES leads on a length
scale of at least 10~$\mu$m (elastic mean free path or a longer
relevant length scale). It's tempting to interpret the drag-type
experiments in such nanostructures in terms of purely one- or
zero-dimensional physics. However, the surrounding leads can
sometimes play a key role, as it is the case for an
acoustic-phonon-based interaction between the two nano-circuits
~\cite{counterflow,DQDratchet,schinner,gasser,gasser_nanotech}.
One way to verify the importance of the leads is to apply a small
perpendicular magnetic field ($B$), which modifies charge
transport in 2DES's thanks to a Lorentz force. At the same time,
the impact of small $B$ on nanostructures can often be reduced to
a trivial magnetic-subbands depopulation, as, e.g., in a quantum
point contact (QPC)~\cite{Fertig_Halperin}. Application of $B>1$T
was relevant in refs.~\cite{yamamoto,onac}, however the role
played by the 2DES leads in those experiments have not been
carefully analyzed.

In this paper, we investigate the influence of $B$ on the
acoustic-phonon-based interaction between the leads of two
coplanar isolated QPCs in a drag-type experiment. At $B=0$ the
counterflow effect in such a system has been recently
observed~\cite{counterflow}, with currents in the drive-circuit
($I_{\rm DRIVE}$) and detector-circuit ($I_{\rm DET}$) flowing in
the opposite directions. We find that in moderate $B$ the
direction of $I_{\rm DET}$ is insensitive to that of $I_{\rm
DRIVE}$, and, instead, determined solely by the polarity of $B$.
This is interpreted in terms of a non-equilibrium skipping orbits
transport along the edges of the 2DESs. Our observations (i)
demonstrate that the role played by the leads can be decisive in
drag-type experiments in magnetic field and (ii) provide a
possible way to verify its importance.

\begin{figure}
\begin{center}
\includegraphics[width=0.6\columnwidth]{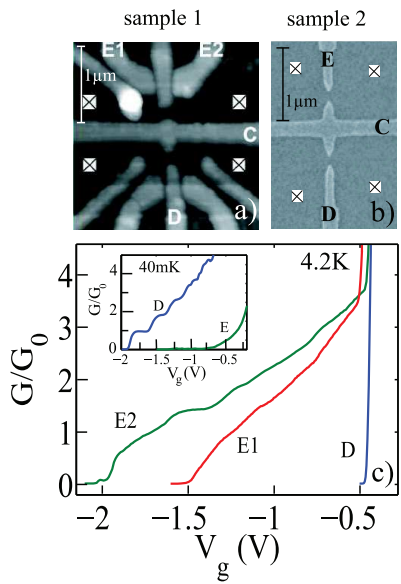}
\end{center}
\caption{(a) Atomic-force micrograph of the split-gate
nanostructure on the surface of sample 1. (b) Scanning electron
micrograph of the nanostructure equivalent to that of sample 2.
$E,E1,E2$ and $D$, respectively, denote the gates used for the
definition of the drive-QPC and detector-QPC; crosses denote ohmic
contacts. (c) Normalized conductance of the QPCs as a function of
their gate voltages for sample 1 (body) and 2 (inset) at
respective bath temperatures.}\label{fig1}
\end{figure}

Our samples are prepared on a GaAs/AlGaAs heterostucture
containing a 2DES 90~nm below the surface with an electron density
of $2.8\times10^{11}cm^{-2}$ and a low-temperature mobility of
$1.4\times 10^6 cm^{-2}/Vs$. For each of the two samples, a number
of metallic gates is deposited with e-beam lithography on the
surface of the crystal, as shown in figs.~\ref{fig1}a
and~\ref{fig1}b. Applying a negative voltage on the central gate
(C), one creates two electrically isolated circuits, separated by
a distance of several hundreds nanometers. In each circuit, a QPC
is defined via negative biasing a metallic gate on a corresponding
side of the gate C. An external dc bias voltage, $V_{\rm DRIVE}$,
is applied to one lead of the drive-QPC, while its second lead is
grounded. A current $I_{\rm DET}$ generated in the circuit of the
detector-QPC is measured with a home-made $I$-$V$ converter
(conversion ratio $10^{-9}$~A/V) connected to one of its leads.
The other lead of the detector-QPC is kept at an input offset
potential of the $I$-$V$ converter ($<10\mu V$), to maintain the
detector-circuit unbiased. In both circuits, a positive sign of
the current is chosen for electrons flowing to the left. Our
results were obtained with both ac and dc measurements. For ac
measurements, $V_{\rm DRIVE}$ was modulated with a small voltage
of $100\mu V$ at frequencies in the range 7-33~Hz. The real phase
ac component of $I_{\rm DET}$ was measured with a lock-in. The
experiments were performed in a liquid $^4$He cryostat at a
temperature of 4.2 K (sample 1) and a $^3$He/$^4$He dilution
refrigerator at a base temperature of 40~mK (sample 2). Exchanging
either detector- and drive-circuit or the bias and ground leads of
the drive-QPC yields qualitatively the same results. This is also
true if in sample 1 either gate E1 or E2 are used to define the
drive-QPC (data given below are obtained with E2). The origin of
the measured detector signals is only consistent with the
acoustic-phonon-based interaction. The corresponding discussion of
Refs.~\cite{DQDratchet,counterflow} takes into account the
interaction bandwidth, the strongly nonlinear response, the
insensitivity to physical distance between the QPCs and the
conservation laws. Apparent unimportance of a two orders of
magnitude temperature variation as well as $B$-dependencies
observed here also fit into the phononic scenario.

Fig.~\ref{fig1}c displays linear-response conductance for each of
the QPCs used in our experiments. The conductance is zero at low
gate voltages (QPC pinched-off) and goes up with increasing gate
voltage. A clear conductance quantization is visible only for one
of the QPCs studied (D for sample 2), whereas a poor quality of
others and/or high temperature (for sample 1) prevent its
observation. Note, though, that the key ingredient for
thermoelectric-like effects is the energy dependence of the
detector-QPC transparency~\cite{counterflow,kamenev}, rather than
the conductance quantization. Throughout the paper, the
detector-QPC is kept near the pinch-off with resistance
$\sim10^2~k\Omega$. The drive-QPC is either (i) relatively open or
(ii) nearly pinched-off. The serial resistance of the ohmic
contacts is negligible in all cases except for the open drive-QPC
regime in sample 2. In the last case, the voltage drop
corresponding to $B=0$ resistance of the ohmic contacts was
subtracted from $V_{\rm DRIVE}$. First, we concentrate on the open
drive-QPC regime with $R_{\rm DRIVE}\sim10k\Omega$.

\begin{figure}
\begin{center}
\includegraphics[width=0.8\columnwidth]{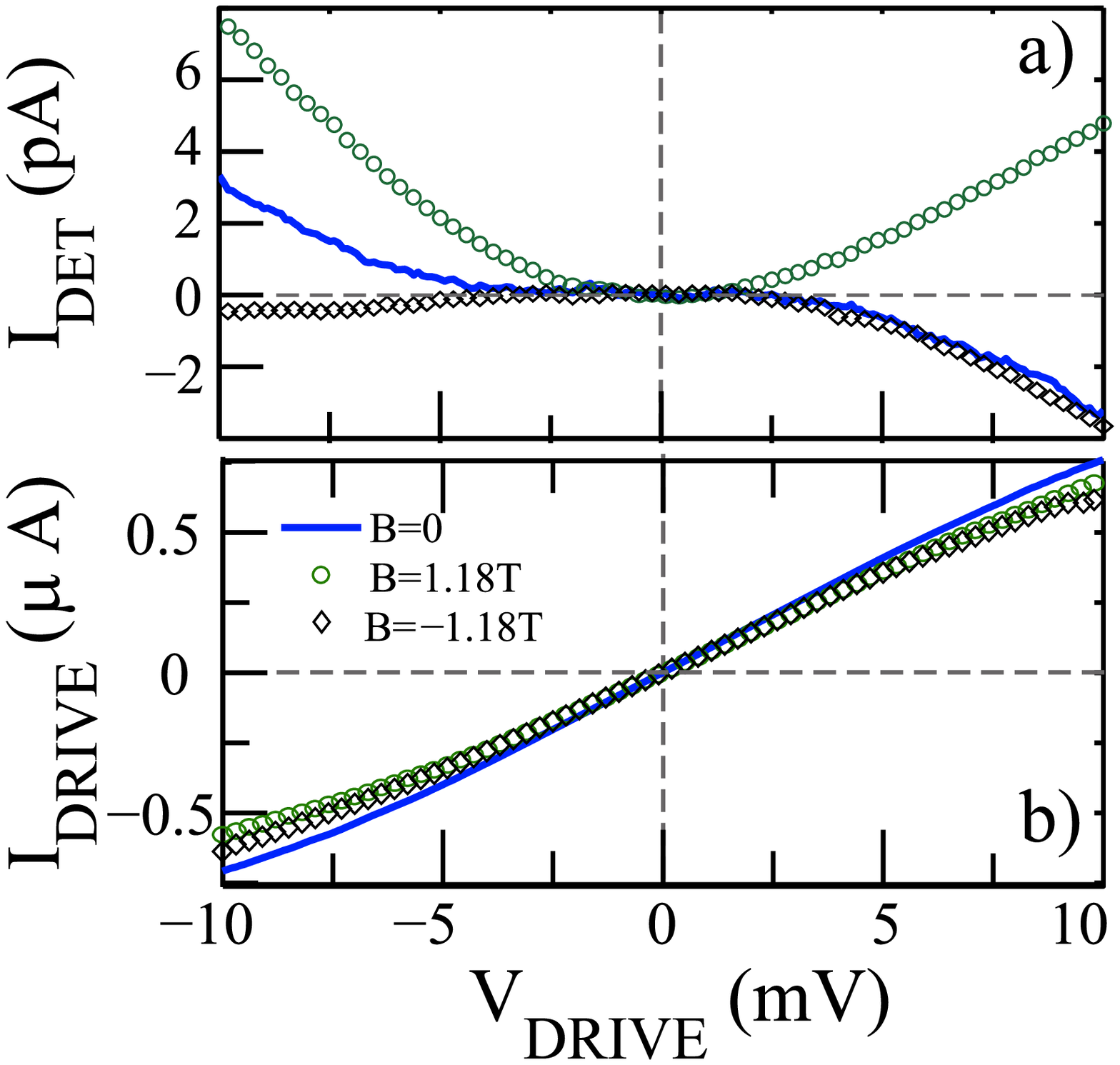}
\end{center}
\caption{Results for sample 1 at 4.2~K. (a) Detector current at
$B=0$ (solid line) and in moderate magnetic fields (symbols, same
legend as in panel b) as a function of drive bias. (b) $I-V$
curves of the drive-QPC taken simultaneously with the data in (a).
}\label{fig2}
\end{figure}

At $B=0$ we reproduce previous results obtained on a different
sample~\cite{counterflow}. The current induced in the detector-QPC
at $B=0$ is shown in fig.~\ref{fig2}a as a function of $V_{\rm
DRIVE}$ by a solid line (sample 1). $I_{\rm DET}$ is a
threshold-like function of $V_{\rm DRIVE}$, changing its sign on
reversal of $V_{\rm DRIVE}$. For $V_{\rm DRIVE}\gtrsim$3~mV a
finite $I_{\rm DET}$ is observed, directed opposite to $I_{\rm
DRIVE}$ (referred to as a counterflow effect
in~\cite{counterflow}). In relatively high $|B|\sim1$~T applied
perpendicular to the 2DES the situation is changed dramatically.
The sign of $I_{\rm DET}$ is independent of that of $I_{\rm
DRIVE}$ (symbols in fig.~\ref{fig2}). Instead, it is determined by
the sign of $B$ ($B>0$ corresponds to magnetic field pointing
downward into the crystal). Below we refer to this regime as a
$B$-driven regime. At the same time, no effect of $B$ is seen in
the drive-QPC $I$-$V$ curve, apart from a minor increase of its
resistance (fig.~\ref{fig2}b). The absolute value of $I_{\rm DET}$
in magnetic field depends on the polarity of $V_{\rm DRIVE}$
(fig.~\ref{fig2}a), which is discussed below.

\begin{figure}
\begin{center}
\includegraphics[width=1\columnwidth]{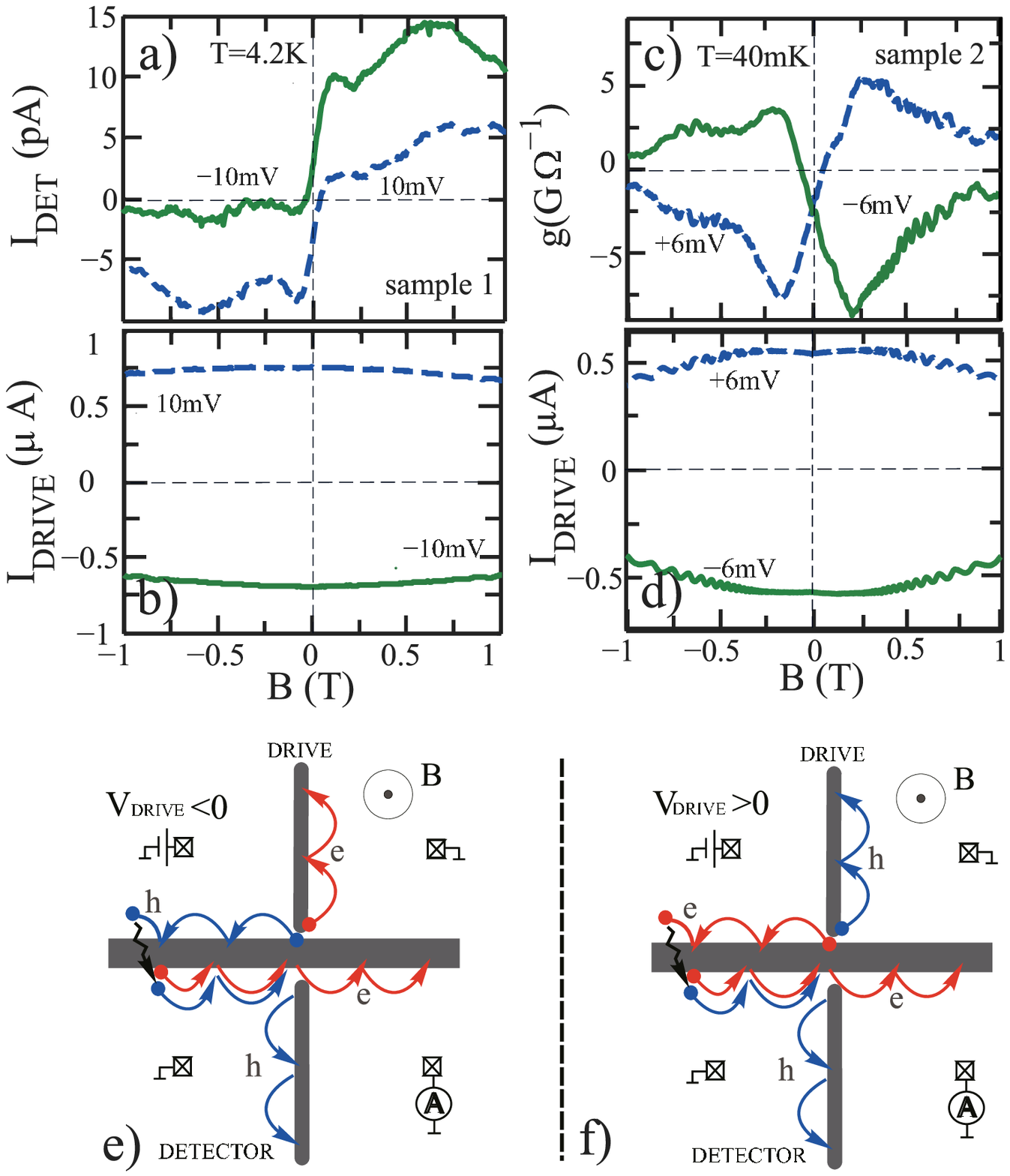}
\end{center}
\caption{Open regime with $R_{\rm DRIVE}\sim10k\Omega$. (a),(b)
Detector (a) and drive (b) currents at fixed $V_{\rm
DRIVE}=\pm10$mV as a function of magnetic field $-1{\rm T}\leq
B\leq1{\rm T}$ for sample 1. (c),(d) Derivative $g$ of the
detector signal (c)  and drive current (d) at fixed $V_{\rm
DRIVE}=\pm6$mV as a function of magnetic field $-1{\rm T}\leq
B\leq1{\rm T}$ for sample 2. (e),(f) Quasi-classical trajectories
of non-equilibrium carriers skipping along the electrostatic edges
of the 2DES nearby the drive-QPC and detector-QPC. Sketches are
drawn for $B<0$ and $V_{\rm DRIVE}<0/>0$ (e)/(f). Arrows indicate
the direction of electrons group velocity near the edge.
Schematics of the measurement connections is also shown, with
ohmic contacts represented by crossed squares.}\label{fig3}
\end{figure}

In fig.~\ref{fig3}a we show how the detector current evolves as a
function of $B$ in sample 1 at fixed $V_{\rm DRIVE}=\pm10$~mV. A
transition to the $B$-driven regime occurs at small fields. Here,
$I_{\rm DET}$ steeply increases, linear in $B$, and changes the
sign when $B$ is swept from -0.1~T to +0.1~T. Outside this
interval, in the $B$-driven regime, no strong B-dependence is
observed. The same qualitative behavior is found in sample 2 with
a differential ac-measurement. In fig.~\ref{fig3}c the measured
derivative $g\equiv dI_{\rm DET}/dV_{\rm DRIVE}$ is plotted for
$V_{\rm DRIVE}=\pm6$~mV. Note, that, e.g., a positive $I_{\rm
DET}$ at a negative $V_{\rm DRIVE}$ results in $g<0$. At $B=0$ the
counterflow effect is again reproduced with $g<0$ regardless the
sign of $V_{\rm DRIVE}$. With increasing $|B|$ a transition to the
$B$-driven behavior occurs in fig.~\ref{fig3}c, which manifests
itself as a sign change of $g$ in response to $V_{\rm DRIVE}$
reversal. As seen from figs.~\ref{fig3}b and~\ref{fig3}d, for both
samples, the simultaneously measured $I_{\rm DRIVE}$ is a
featureless weakly-parabolic function of $B$ with a small
oscillatory contribution associated with the Shubnikov-de-Haas
effect at low temperature (d).

At $B=0$, the counterflow effect has been explained in a
thermoelectric analogy~\cite{counterflow}. The associated spatial
asymmetry of the energy flow from the drive-circuit occurs in the
nonlinear regime and is determined by the direction of the current
$I_{\rm DRIVE}$. The thermoelectric analogy is also applicable to
the $B$-driven regime, where the direction of the $I_{\rm DRIVE}$
is irrelevant to that of the $I_{\rm DET}$. The origin of the
asymmetry in this case is related to a cyclotron motion  of charge
carriers in perpendicular magnetic field. In a quasi-classical
picture, bulk 2D electrons circle around cyclotron orbits of a
radius $R_C\propto|B|^{-1}$ thanks to the Lorentz force. In
sufficiently high $B$, such that $R_C<L$, where $L\approx11~\mu$m
is the elastic mean free path set by a disorder in our samples,
the electrons' trajectories are localized within regions of size
smaller than $L$. This is not the case for carriers incident on or
leaving from a QPC, whose cyclotron circles cross the
electrostatic edge of the 2DES. These carriers exhibit a specular
reflection at the edge and follow skipping orbit trajectories.

In figs.~\ref{fig3}e and~\ref{fig3}f the motion of non-equilibrium
carriers next to the QPCs is sketched for $B<0$ and different
signs of $V_{\rm DRIVE}$. Here, $e$ and $h$ stand for
non-equilibrium electrons above and non-equilibrium holes (i.e.
unoccupied electron states) below the Fermi energy, respectively.
In the (upper) drive-circuit the non-equilibrium $e$ and $h$ exit
on different sides of the drive-QPC, as determined by the sign of
$V_{\rm DRIVE}$. In the vicinity of the constriction, the carriers
in the drive-circuit follow two quasi-classical trajectories
skipping along the respective edges of the 2DES electrostatically
defined by the split-gates. One of them skips along the
(horizontal) central gate C and the other along the (vertical)
drive-QPC gate. The non-equilibrium carriers moving along the gate
C ($e/h$ for $V_{\rm DRIVE}>0/<0$) are most efficient in
transferring energy to the adjacent lead of the detector-circuit
(zig-zag arrows in figs.~\ref{fig3}e and~\ref{fig3}f). As
discussed in Ref.~\cite{schinner}, such energy transfer is likely
to be mediated by acoustic phonons which are reabsorbed in the
detector within $\sim1~\mu$m. As a result, non-equilibrium $e-h$
pairs are excited in the detector, moving along the edge towards
the detector-QPC. Note, that non-equilibrium population of the
skipping orbit states can also occur indirectly, via excitation of
$e$ and $h$ in the bulk 2DES, with their subsequent diffusion to
the edge. On the average, more non-equilibrium electrons traverse
the detector-QPC, because of the energy dependence of the QPC
transparency~\cite{counterflow}, which results in the measured
$I_{\rm DET}$. As follows from figs.~\ref{fig3}e and~\ref{fig3}f,
$I_{\rm DET}<0$ for $B<0$, independent of the direction of $I_{\rm
DRIVE}$. Similar sketches would be obtained for $B>0$ via
mirroring about the vertical axis and simultaneously changing the
sign of $V_{\rm DRIVE}$, which explains the detector current
reversal: $I_{\rm DET}>0$ for $B>0$.

The reversal of $V_{\rm DRIVE}$ corresponds to changing the type
of non-equilibrium carriers skipping along the gate C in the
drive-circuit (compare figs.~\ref{fig3}e,~\ref{fig3}f). The
fraction of the total energy (later dissipated in the leads as
Joule heat) injected with $e$ in one lead of the drive-QPC is
higher than the fraction injected with $h$ in the other lead. This
comes from the positive energy-dependence of the QPC
transparency~\cite{counterflow}. In the $B$-driven regime, this
$e-h$ asymmetry manifests itself as a higher absolute value of
$I_{\rm DET}$ and $g$ for the $e$-sign of $V_{\rm DRIVE}$ (a
slight $V_{\rm DRIVE}$-asymmetry of the total Joule heat inferred
from figs.~\ref{fig3}b,~\ref{fig3}d has a minor effect here).
Correspondingly, in figs.~\ref{fig3}a and~\ref{fig3}c we find the
detector response stronger for $V_{\rm DRIVE}>0$ ($<0$) at $B<0$
($>0$). This behavior is more pronounced in sample 1
(figs.~\ref{fig3}a), which is likely caused by a stronger
energy-dependence of the drive-QPC transparency in this case.

The regions of the 2DESs where emission/reabsorption of the
acoustic phonons is efficient are different in the counterflow
regime ($B=0$) and in the $B$-driven regime. The transition to the
$B$-driven regime occurs at $|B|\sim50$mT (figs.~\ref{fig3}a,c),
which corresponds to $R_C\approx1.7~\mu$m~\cite{remark}. This
length-scale is consistent with that found for acoustic
phonon-based interaction between the coplanar
2DESs~\cite{schinner}. Such a small length-scale supports the
interpretation in terms of ballistic skipping orbits. In contrast,
a qualitatively similar classical picture of hot-spots in a 2DES
in strong $B$ would give a typical scale at least an order of
magnitude larger~\cite{dietsche}.

The lowest order linear behavior $\delta I_{\rm DET},\delta g
\propto B$ in figs.~\ref{fig3}a and~\ref{fig3}c indicates that
breaking of the time-reversal symmetry in magnetic field is
important, which is indeed the case for chiral transport via
skipping orbits. The same is true, e.g., for the Hall effect or
transverse thermopower (Nernst-Ettingshausen (NE) effect). In
fact, the data of fig.~\ref{fig3}a resembles the behavior of the
NE coefficient in a bulk 2DES in perpendicular
fields~\cite{fletcher}. The similarity suggests that the energy
flow from the drive-circuit is analogous to a temperature
gradient. Still, there is an important difference between the
experiments. The NE effect comes from the energy dependence of the
momentum relaxation time (diffusion contribution) and asymmetry in
electron-phonon interaction (phonon drag
contribution)~\cite{fletcher}. In our case, the key ingredient is
likely to be the energy dependence of the detector-QPC
transparency~\cite{counterflow}.
\begin{figure}
\begin{center}
\includegraphics[width=1\columnwidth]{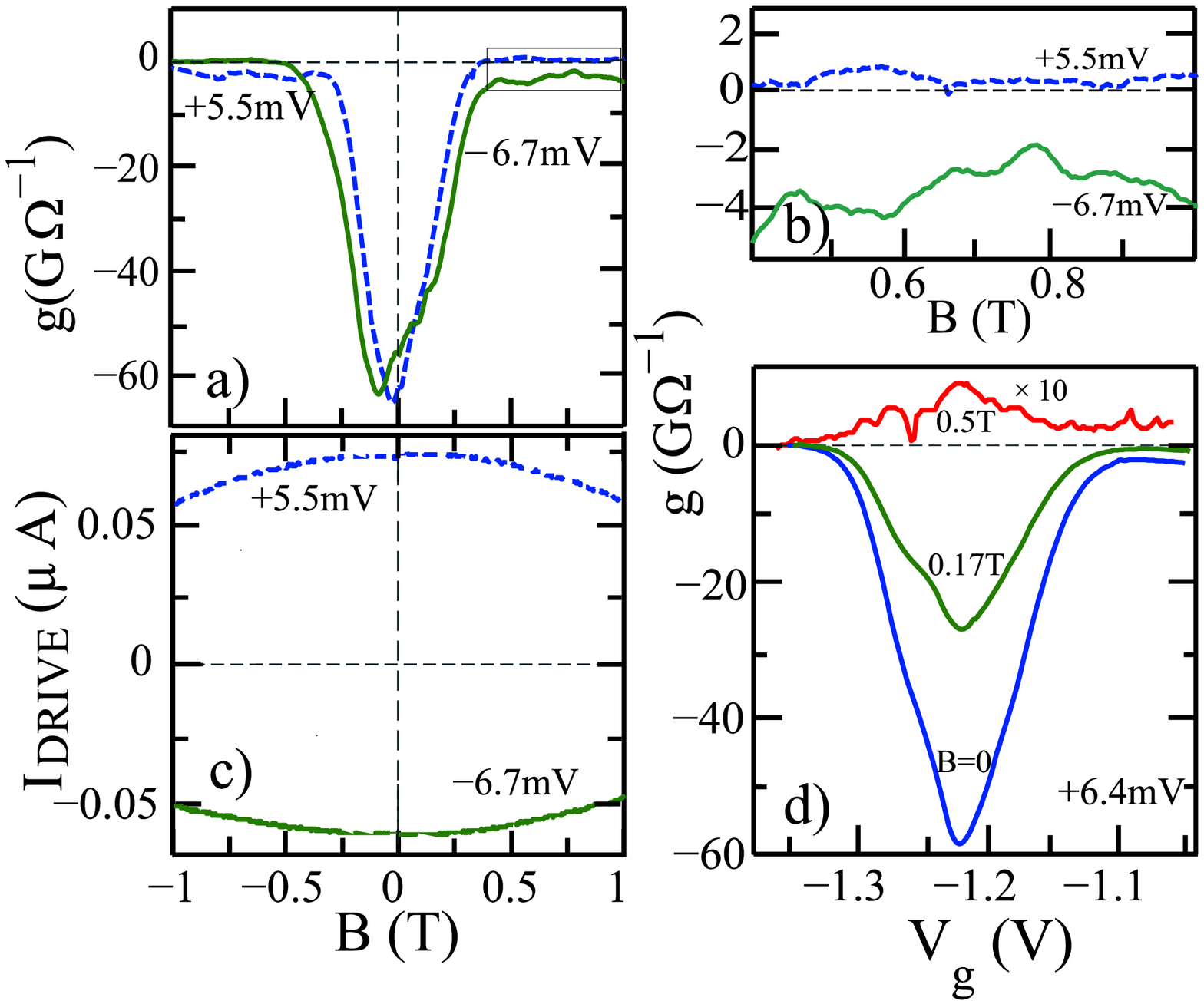}
\end{center}
\caption{Regime of the nearly pinched-off drive-QPC, sample 2,
bath temperature of 40mK. (a),(c) Derivative $g$ of the detector
signal for two values of $V_{\rm DRIVE}$ as a function of $B$ and
the corresponding traces for $I_{\rm DRIVE}$. (b) Blown-up $B>0$
region of (a), where a $B$-driven behavior is seen, similar to
fig.~\ref{fig3}c. (d) Derivative $g$ of the detector signal at
fixed $V_{\rm DRIVE}$ and several values of $B$ as a function of
the gate voltage, controlling the drive-QPC transparency. In
(a)-(c) the linear-response resistance of the drive-QPC is $R_{\rm
DRIVE}\sim80k\Omega$.}\label{fig4}
\end{figure}

Next, we briefly outline the results for the drive-QPC tuned near
pinch-off (and the same settings for the detector-QPC).
Fig.~\ref{fig4}d shows the derivative $g$ plotted versus the
drive-QPC gate voltage near its pinch-off for three $B$-field
values. At $B=0$, the counterflow is observed and $g$ exhibits a
pronounced extremum. The absolute value is nearly two orders of
magnitude larger compared to the open regime (fig.~\ref{fig3}c),
as reported earlier~\cite{DQDratchet,counterflow}.
Fig.~\ref{fig4}d shows that an application of a moderate
$|B|\sim0.5$~T drastically suppresses the signal. The
$B$-dependence is plotted in fig.~\ref{fig4}a for two values of
$V_{\rm DRIVE}$ (here, the gate voltage is tuned to the $B=0$
extremum in fig.~\ref{fig4}d). Near zero field, the detector
signal is roughly even in $B$ and no steep $B$-dependence is
observed, in sharp contrast to the case of open drive-QPC
(fig.~\ref{fig3}c). Transition to the $B$-driven regime is found
only for $|B|\gtrsim0.5$~T (figs.~\ref{fig4}a,~\ref{fig4}b
and~\ref{fig4}d) with the detector signal comparable to the case
of open drive-QPC. Note, that the impact of the $e-h$ asymmetry
discussed above is much stronger in fig.~\ref{fig4}b than in
fig.~\ref{fig3}c, which is a result of hot-$e$ injection across
the drive-QPC near pinch-off~\cite{palevski}. As shown in
fig.~\ref{fig4}c, all these observations are accompanied by a
little $B$-dependence of $I_{\rm DRIVE}$. The enhanced driving
efficiency for the almost pinched-off drive-QPC and its low-$B$
behavior is hard to ascribe to the phonon emission deep in the
2DES leads, unlike for the open drive-QPC. Possibly, the energy
relaxation of hot-$e$ close ($\ll1\mu m$) to a pinched-off QPC is
relevant here. Although this is consistent with experiments on
transverse focusing of hot-$e$~\cite{williamsonSSc}, we failed to
understand the gate-voltage and $B$-field
dependencies~\cite{DQDratchet}.

The above discussed phononic effects can be considered as quite
general to coupled nano-circuits in magnetic field. The conditions
of the present experiment (a relatively small $B$ and a strongly
non-linear regime) are different from those of, say,
refs.~\cite{yamamoto,onac,Heiblum_dephase}, preventing a direct
comparison. Still, we find it surprising, that the
acoustic-phonon-based interaction have not been carefully ruled
out in those experiments.

In summary, we have explored the influence of a perpendicular $B$
on acoustic-phonon-based interaction between coplanar
nano-circuits in a 2DES. The primary effect of the $B$ is to
modify the motion of non-equilibrium carriers in the 2DES leads
from a ballistic bulk transport to a skipping-orbit edge
transport. As a result, a new $B$-driven regime is found in
moderate $B$, where a spatial asymmetry of phonon
emission/reabsorption is determined solely by field polarity. The
observations suggest a possible way to verify the importance of
the 2DES related effects in some drag-type experiments.

We acknowledge discussions with R.V.~Parfen'ev, D.~Bagretz,
A.~Kamenev, D.V.~Shovkun, V.T.~Dolgopolov, A.A.~Shashkin,
A.A.~Zhukov, E.V.~Deviatov, R.~Fletcher. Financial support by
RFBR, the grant MK-3470.2009.2, and the German Excellence
Initiative via the "Nanosystems Initiative Munich (NIM)" is
gratefully acknowledged. VSK acknowledges support from the
Humboldt Foundation and the Russian Science Support Foundation.


\begin{thebibliography}{20}
\bibitem{remark} A length-scale estimate is quite rough, as
follows from a factor of 2 different $B$-scales of the linear
regions in figs.~\ref{fig3}a and ~\ref{fig3}c near zero magnetic
field. The origin of the discrepancy was not studied.
\end{thebibliography}
\end{document}